\documentclass[compsoc,conference,a4paper,10pt,times]{IEEEtran}
\IEEEoverridecommandlockouts
\usepackage{cite}
\usepackage{amsmath,amssymb,amsfonts}
\usepackage{textcomp}
\usepackage{bmpsize}
\usepackage{xcolor}
\usepackage{lipsum}
\usepackage[colorlinks=true,urlcolor=black]{hyperref}
\def\BibTeX{{\rm B\kern-.05em{\sc i\kern-.025em b}\kern-.08em
    T\kern-.1667em\lower.7ex\hbox{E}\kern-.125emX}}

\usepackage{graphicx}
\usepackage{url}
\usepackage{color}    
\usepackage[ruled,vlined]{algorithm2e}
\usepackage{xspace}
\usepackage{msc}
\usepackage{listings}
\usepackage{paralist}
\usepackage{chemarrow}
\usepackage{multirow}
\usepackage[font={small},skip=4pt,belowskip=5pt]{caption}
\newcommand{\subparagraph}{} 
\usepackage[compact]{titlesec}

\setdefaultleftmargin{1em}{1em}{}{}{}{}
\makeatletter
\g@addto@macro\normalsize{%
  \setlength\abovedisplayskip{4pt}
  \setlength\belowdisplayskip{4pt}
  \setlength\abovedisplayshortskip{4pt}
  \setlength\belowdisplayshortskip{4pt}
}
\makeatother

\newcommand{\name}{SmartCert\xspace}
\newcommand{\myparagraph}[1]{\noindent\textbf{#1}~\xspace}

\renewcommand{\tablename}{Tab.}

\renewcommand{\figurename}{Fig.}

\begin{document}
\title{\name: Redesigning Digital Certificates with Smart Contracts}
\author{Pawel Szalachowski\\
    SUTD, Singapore
}


\maketitle

%
\begin{abstract}
    Digital certificates are usually essential for achieving secure
    communication in open and untrusted environments like the Internet.  The
    Transport Layer Security (TLS) protocol and its public-key infrastructure
    (PKI) are widely used in today's Internet to achieve such a secure
    communication.  Validating domain ownership by trusted certification
    authorities (CAs) is a critical step in issuing digital certificates, as the
    security of all subsequent connections depends on this validation.
    Unfortunately, this validation process provides 
    a poor security level, and  although many improvements have been
    proposed to the date, they suffer from security and deployability issues.
    In this work, we present \name, a novel approach to redesign and
    improve the properties of digital certificates.  \name is empowered by smart
    contracts and thanks to this technology \name can provide benefits of the
    existing PKI enhancements as well as new desired functionalities and
    features.  A certificate in \name conveys more detailed information about
    its validation state which is constantly changing but only with respect to
    the specified smart contract code and individual domain policies.  CAs
    issuing and updating certificates are kept accountable and their actions are
    transparent and monitored by the code.  We also present the implementation
    and evaluation of \name, and our results indicate that the system is
    efficient and deployable as today.
\end{abstract}

%

\section{Introduction}
\label{sec:intro}
Secure Internet communication often relies on digital certificates that contain
bindings between subjects (usually domain names) and their public keys.  It is
security-critical that these bindings are correct; otherwise, an adversary
able to obtain a certificate with an incorrect binding could impersonate the
attacked subject.  In the TLS PKI~\cite{rfc5280}, 
CAs are obligated to validate the correctness of
these bindings and subsequently are allowed to issue certificates stating it.
Unfortunately, the validation process provides a poor security level
since it is conducted only once 
for the certificate lifetime~\cite{bhargavan2017formal}.  Moreover, the TLS PKI
provides the weakest-link security and an adversary compromising a single CA
(out of hundreds~\cite{perl2014you}) can issue a certificate
for any domain.  A certificate also has to be accompanied
with its revocation status (as it can be prematurely
invalidated) but designing a sustainable and effective
revocation system turned out to be a huge challenge~\cite{liu2015emc}. 

There were many approaches proposed to solve or mitigate these issues.  Extended
Validation (EV) certificates~\cite{EVguidelines} try to introduce more secure
public-key validation by requiring a requesting entity to prove
that it has a legal control over the domain.  Unfortunately, EV
certificates are unpopular~\cite{ouvrier2017characterizing}, their security
benefits are questionable~\cite{jackson08fgo,EVvalue}, and their support in main
browsers may be deprecated in a near future~\cite{evcab}.
Another approach is notary servers~\cite{wendlandt2008perspectives,marlinspikeconvergence}, which
contact TLS servers and report observed public keys
to a TLS client that can compare its own view with the views of notary server(s)
prior to accepting the TLS connection.
Although properties provided by notary systems are desired in an
environment like the Internet, their deployment issues thwarted their adoption.
Certificate Transparency (CT)~\cite{rfc6962} is a prominent log-based system
which does not improve the public-key validation process but rather aims to make
actions of CAs visible by logging all issued certificates.  There have been
many proposals for a secure and efficient revocation
system~\cite{rfc5280,crlset,rivest1998can}; however, despite them, the current
TLS PKI still cannot offer this basic functionality~\cite{liu2015emc}.  The
domain expressiveness in the TLS PKI is also unsatisfactory. Domains can use
CAA~\cite{rfc6698} or DANE~\cite{rfc6844} to specify
simple policies, like their trusted CAs; however, these systems rely
upon the DNS/DNSSEC infrastructure causing inefficiencies, and low security
and deployability~\cite{scheitle2018first}.

One important observation is that in the current TLS PKI certificates are
static; thus are: \textit{a)} inflexible in terms of adding and handling new
features, and \textit{b)} decoupled from the information about their validity
(like policies or a revocation status).  Such a design causes serious
issues, as the security and availability of the certificate
validation process are underpinned by multiple separate infrastructures (e.g., for
revocation or logging).  
In this paper, we present \name, a framework where certificates are empowered by
smart contracts. This design choice gives us an opportunity to rethink and
redesign how digital certificates can be maintained and expressed in order to
improve their security and flexibility. In \name, certificates are dynamic
objects carrying their validation states, which are able to evolve according to
pre-defined rules and individual domain policies. These rules and policies are
self-enforced by a code; therefore, the trust placed in CAs can be significantly
minimized.  We present \name in the context of improving dominant
domain-validated certificates but the framework can be extended to other
validation policies easily.  \name combines advantages of previous systems and
with ``history-rich'' certificates provides new interesting properties
beneficial for clients and domains.  \name is deployable with the legacy
TLS infrastructure, requires small changes on involved parties and does not
require any new infrastructure to be deployed.  We implemented and evaluated
\name, and our results indicate its efficiency and practicality.

\section{Background}
\label{sec:pre}

\subsection{Notation}
We use the following notation.
\begin{inparadesc}
    \item[$H(.)$] is a cryptographic hash function\textbf{;}\quad
    \item[$\|$] is the string concatenation\textbf{;}\quad
    \item[$\{0,1\}^n$] is a set of all $n$-bit long strings\textbf{;}\quad
    \item[$r\xleftarrow{R}A$] denotes that $r$ is an element randomly selected
        from the set $A$\textbf{;}\quad
    \item[$\mathit{Sign_{sk}(m)}$] for
        a private key $sk$ and a message to sign $m$, creates the
        corresponding signature $\sigma$\textbf{;}\quad
    \item[$\mathit{SigVrfy_{pk}(m, \sigma)}$] returns
        \textit{True} if the message $m$ matches the signature $\sigma$ under
        the public key $pk$, \textit{False} otherwise\textbf{;}\quad
    \item[$\mathit{time()}$] returns the current timestamp.
\end{inparadesc}

\subsection{Transport Layer Security}
\label{sec:pre:tls}
The TLS protocol~\cite{rfc5246}
provides security for the client-server model
and can be used to secure any application-layer protocol; however, its main use
case on the Internet is HTTPS, i.e., running the HTTP protocol on the top of
TLS. (We focus on TLS 1.2 as it is currently
recommended and widely deployed, but \name can be easily combined with
TLS 1.3 as we discuss in \autoref{pre:practical:tls13}.) TLS consists of
multiple subprotocols, one of which is the TLS handshake protocol. TLS handshake
initiates a TLS connection and is designed to negotiate a shared symmetric key
between a client and a server. During an execution of this protocol, the
communicating parties decide to accept or reject the connection.  There are many
ways this shared key can be negotiated; however, methods
based on the elliptic curve ephemeral Diffie-Hellman (DH)
protocol~\cite{blake1998authenticated} are recommended and used by default by
modern TLS implementations~\cite{whytls13,ssllabs}.

\newcommand\Ra[1]{$\autorightarrow{\small\raisebox{-1.0ex}[0pt][-1.5ex]{#1}}{} $}
\newcommand\La[1]{$\autoleftarrow{\small\raisebox{-1.0ex}[0pt][-1.5ex]{#1}}{} $}
\newcommand\CHrnd{\texttt{CH}\textit{.r}\xspace}
\newcommand\SHrnd{\texttt{SH}\textit{.r}\xspace}

\begin{figure}[b!]
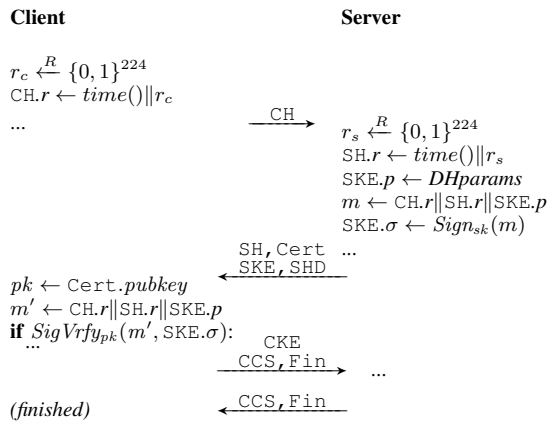

\centering
\scalebox{.875}{
\small
    \begin{tabular}{@{}l@{\hspace{-0.25cm}}@{}c@{}@{\hspace{-0.1cm}}l@{}}
    \textbf{Client} & & \textbf{Server} \\\\
    $r_c\xleftarrow{R} \{0,1\}^{224}$  &  &  \\
        $\CHrnd\leftarrow time()\|r_c$  &  &  \\
\vspace{-0.5cm}
    ... & \Ra{\texttt{CH}}  &  \\
    & & $r_s\xleftarrow{R} \{0,1\}^{224}$  \\
        & & $\SHrnd\leftarrow time()\|r_s$  \\
    & & $\texttt{SKE}\textit{.p}\leftarrow \textit{DHparams}$\\
    & & $m\leftarrow \CHrnd\|\SHrnd\|\texttt{SKE}\textit{.p}$ \\
    & & $\texttt{SKE}.\sigma\leftarrow \mathit{Sign_{sk}}(m)$  \\
    & \texttt{\small SH,Cert}& ...  \\
\vspace{-0.4cm}
    & \La{\texttt{SKE,SHD}} &  \\
    $\mathit{pk\leftarrow\texttt{Cert}.pubkey}$ & & \\
    $m'\leftarrow \CHrnd\|\SHrnd\|\texttt{SKE}\textit{.p}$ & & \\
\vspace{-0.2cm}
    \textbf{if} $\mathit{SigVrfy_{pk}(m', \texttt{SKE}.\sigma)}$:& & \\
    \ \ ... & \texttt{\small CKE} & \\
    & \Ra{\texttt{CCS,Fin}} & \ \ \ \ ...  \\
        \textit{(finished)} & \La{\texttt{CCS,Fin}} &  \\
\end{tabular}
}
    \caption{A TLS handshake with an ephemeral DH key exchange.}
    \label{fig:tls_handshake}
\end{figure}

The TLS handshake with an ephemeral DH key exchange is presented in
\autoref{fig:tls_handshake}.  It is initiated by the client that
sends a ClientHello message (\texttt{CH}) containing a random value
$\CHrnd$ selected by the client (this value is prepended with the client's
timestamp), a list of the cryptographic methods that the client supports,
and other parameters and extensions (optionally). After receiving the
ClientHello message, the server prepares multiple messages to send them back to
the client.  In particular, the ServerHello message (\texttt{SH}) contains a
random field created analogically, while the ServerKeyExchange message
(\texttt{SKE}) contains authenticated information about the key exchange.  This
information consists of the client's random value (\CHrnd), the server's random
value (\SHrnd), parameters of the DH protocol, and the signature ($\sigma$) that
protects all these values.  The \texttt{SKE} message is sent to the client
together with the server's certificate (that allows the client to verify the
server's signature and identity) and with the ServerHelloDone message
(\texttt{SHD}).  Upon receiving these messages the client performs multiple
verification checks including ensuring that the signature of the \texttt{SKE}
message was signed indeed by the entity from the certificate (it is not
presented in the figure but the certificate is also validated at this stage of
the protocol). The client  computes a shared key and
establishes it with the server via the ClientKeyExchange message
(\texttt{CKE}). Then the parties exchange ChangeCipherSpec and Finish
messages indicating 
that the following communication can be encrypted.

\subsection{TLS PKI and Certificates}
\label{sec:pre:x509}
TLS employs the X.509 PKI~\cite{rfc5280}, where CAs are trusted parties
contacted by entities wishing to obtain X.509v3 certificates~\cite{rfc5280}. In
such a case, a CA is obligated to validate the binding between an entity's name
and a presented public key. If the CA successfully validates such a binding, a
certificate stating so can be issued.  In the TLS PKI, the public-key
validation is conducted once per certificate and is usually automated by
following the trust-on-first-use model over an insecure protocol like DNS,
HTTP, or e-mail~\cite{bhargavan2017formal}. It is called a domain validation
and this approach provides poor security as an adversary able to launch a
man-in-the-middle (MitM) attack even for a
moment~\cite{gavrichenkov2015breaking,birge18bamboozling} can obtain a
certificate (for the attacked entity) which will be valid for several months.
(A sophisticated off-path attack was presented recently~\cite{Brandt:2018:DVM:3243734.3243790}.)

In most TLS applications (like HTTPS), only servers have certificates and usually,
they are identified by their associated DNS domain names.  Clients verify
certificates by checking whether they are issued by CAs from the list of
trusted CAs (this list is provided by software vendors --- operating systems,
browsers, or TLS implementations).  

\subsection{Blockchain and Smart Contracts}
\label{sec:pre:blockchain}
Blockchain platforms started emerging with the advent of
Bitcoin~\cite{nakamoto2008bitcoin}.  Bitcoin is a cryptocurrency which for the
first time introduced the concept of decentralized and permissionless consensus
combined with an immutable data structure. This concept was quickly extended to
more powerful systems that besides monetary transfers offer other features and
functionalities. In particular, smart contract platforms received a lot of
attention, as aside from simple monetary transfers they allow users to specify
arbitrary agreements that can be encoded in high-level programming languages
which these platforms offer.  As of writing this paper,
Ethereum~\cite{buterin2013ethereum} is the most popular and developed smart
contract platform.

Ethereum is an open platform where anyone can make a monetary transfer, submit
its smart contract, interact with other contracts, or become a \textit{miner}
(i.e., a node that is running the underlying consensus protocol).  It
provides its native cryptocurrency --- \textit{ether}.  Ethereum users have
unique addresses (identifiers) that are encoded as hashes of their public keys.
Similarly, every smart contract instance also has a unique address that is
derived as a hash of concatenated its creator's address with a nonce.  To
interact with the platform a user signs and sends a \textit{transaction} which
transfers ether to another account or/and calls a contract's method (specifying the
targeted contract, the method to be called, and its arguments).  Transactions are
collected by miners who execute them and run the consensus protocol to append
new transaction blocks to the blockchain (it is required to keep a
consistent version of its global state).  Callers have to supplement their
transactions by an execution cost called \textit{gas} purchased in ether
from the miner who has appended the transaction to the blockchain. Gas
consumption depends on the operations that the call executes.

\begin{figure}[t!]
    \centering
    \includegraphics[width=0.65\columnwidth]{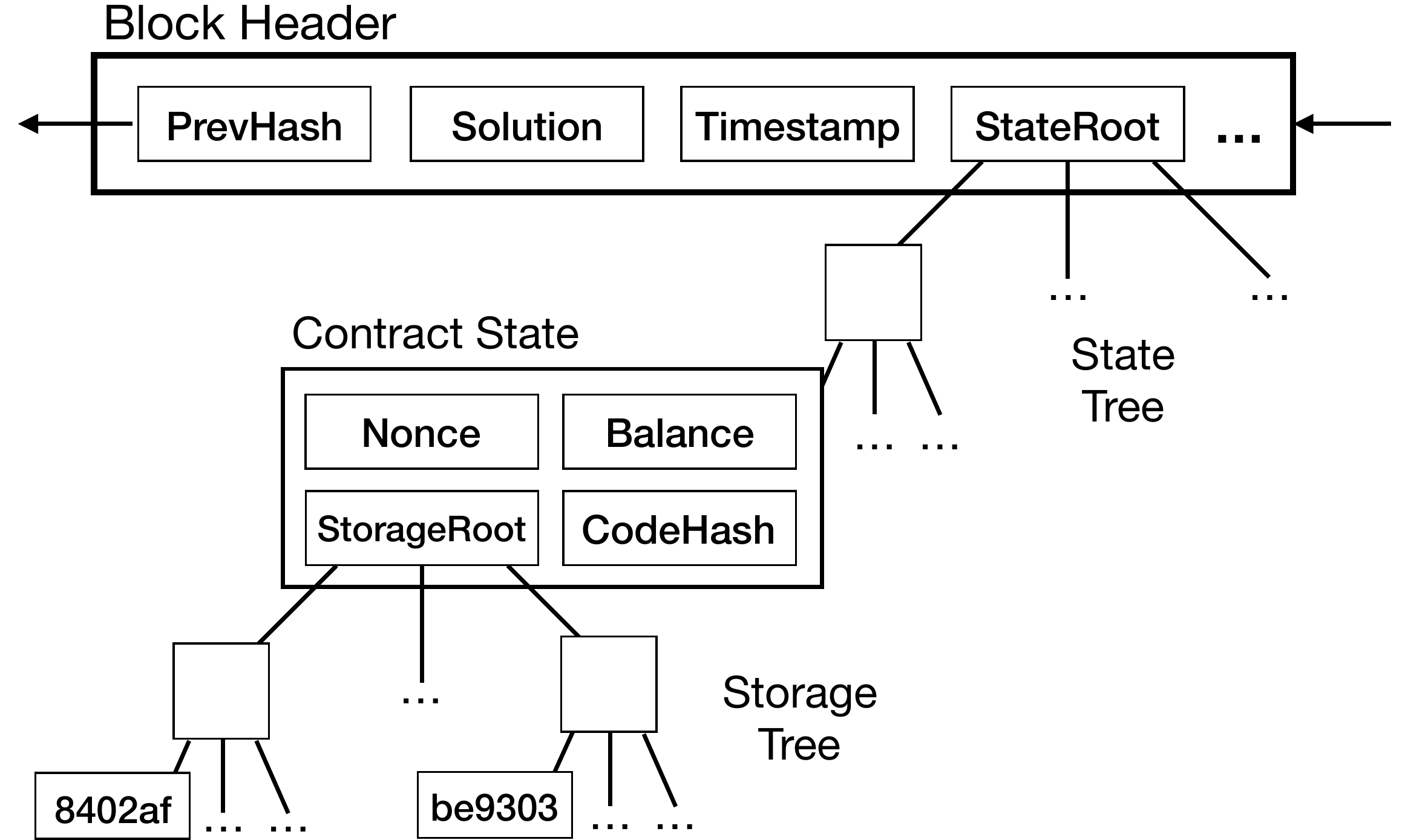}
    \caption{An Ethereum's block header with the associated trees.}
    \label{fig:state}
\end{figure}

Ethereum enables smart contracts to store a persistent storage.  The storage is
maintained by miners and it is part of the global blockchain state.  Each
contract has its state associated, and this state can be accessed through the
\textit{state tree}, which is an instance of the Merkle Patricia Tree (MPT) data
structure~\cite{mpt}.  The state tree is authenticated by including its root in
each \textit{block header}, as depicted in \autoref{fig:state}.  Using the tree,
it is possible to efficiently access a contract's state that includes its nonce,
its balance, the hash of its code, and the root of another MPT 
\textit{storage tree},  including the contract's storage.  Besides
accessing state nodes, an MPT (as a Merkle tree variant) allows to prove that a
given leaf is part of the tree. It is realized with an \textit{MPT proof}, a set
of nodes that allows to rebuild the tree's root.  The size of this set is
logarithmic in the number of MPT leaves.
Thanks to this design, it is possible to prove to \textit{light
clients}~\cite{light_cli} what is the current state (code, balance, and nonce)
or storage of a given smart contract instance.  Light clients are blockchain
clients that store only block headers without storing the entire blockchain.
For instance, to prove that a given storage is part of a contract's state, one
provides a light client with \textit{a)} an MPT proof (of that storage) that
roots in the contract's state node (i.e., \textit{StorageRoot}, see
\autoref{fig:state}), and \textit{b)} an MPT proof (of the state node) that roots
in the state root encoded in a block header (i.e., \textit{StateRoot}, see
\autoref{fig:state}).  With only block headers, a light client can easily
validate these proofs by rebuilding the corresponding MPT roots~\cite{light_cli}.

\section{\name Overview}
\label{sec:overview}
%

\subsection{System Model}
\label{sec:pre:sys_model}

\name introduces the same parties as in the TLS PKI. 
    \textbf{Domain} provides a service via the TLS protocol (e.g., HTTPS). We
        assume that services, and servers that host these services, are
        identified by DNS domain names. (For simple description, we
        use the terms ``domain'' and ``server'' interchangeably.) 
        We assume that domain's servers can interact with the blockchain
        network, obtaining MTP proofs. 
    \textbf{CA} is a trusted party that certifies bindings between identities and
        their public keys. We assume that CAs
        have key pairs that can be used over the blockchain platform, and
        CAs are able to send transactions to the blockchain
        platform.  CAs in \name are similar to TLS PKI CAs, except they
        are not completely trusted and their actions are self-monitored and
        self-enforced by a smart contract code and domain policies.
    \textbf{Client} wishes to contact a service (served by a domain) in a
        secure way.  Clients trust CAs and they are also blockchain light client
        able to obtain the blockchain block headers.

We assume that these parties can interact with
a blockchain platform that supports smart contracts. In particular, we focus on
Ethereum; however, \name can be adjusted to most smart contract platforms
available. 
For demonstration, we assume various adversary models and first, an
MitM adversary is assumed. We assume that MitM
attacks are short-lived; otherwise, a permanent attack would make every
domain-validation scheme ineffective~\cite{wendlandt2008perspectives}.  The goal
of the adversary is to conduct an impersonation attack undetected.  We assume
that the used cryptographic tools are secure and that the adversary cannot
undermine the security properties of the underlying blockchain platform.
Moreover, we assume that CAs can misbehave by not undertaking their duties (or
conducting them incorrectly), but every such misbehavior should be
detectable.  We also consider a stronger adversary who in order to impersonate a
domain can collude with $m$ malicious CAs; however, we assume that the domain
can always prove its identity to $n$ honest CAs, for $n>m$.

\subsection{Design Space}
\myparagraph{TLS PKI Issues}
Certificates in the current TLS PKI do not provide a high level
of security. We discuss some of their design inherent issues that
contribute to this state. 

\begin{figure}[t]
    \centering
    \includegraphics[width=0.9\columnwidth]{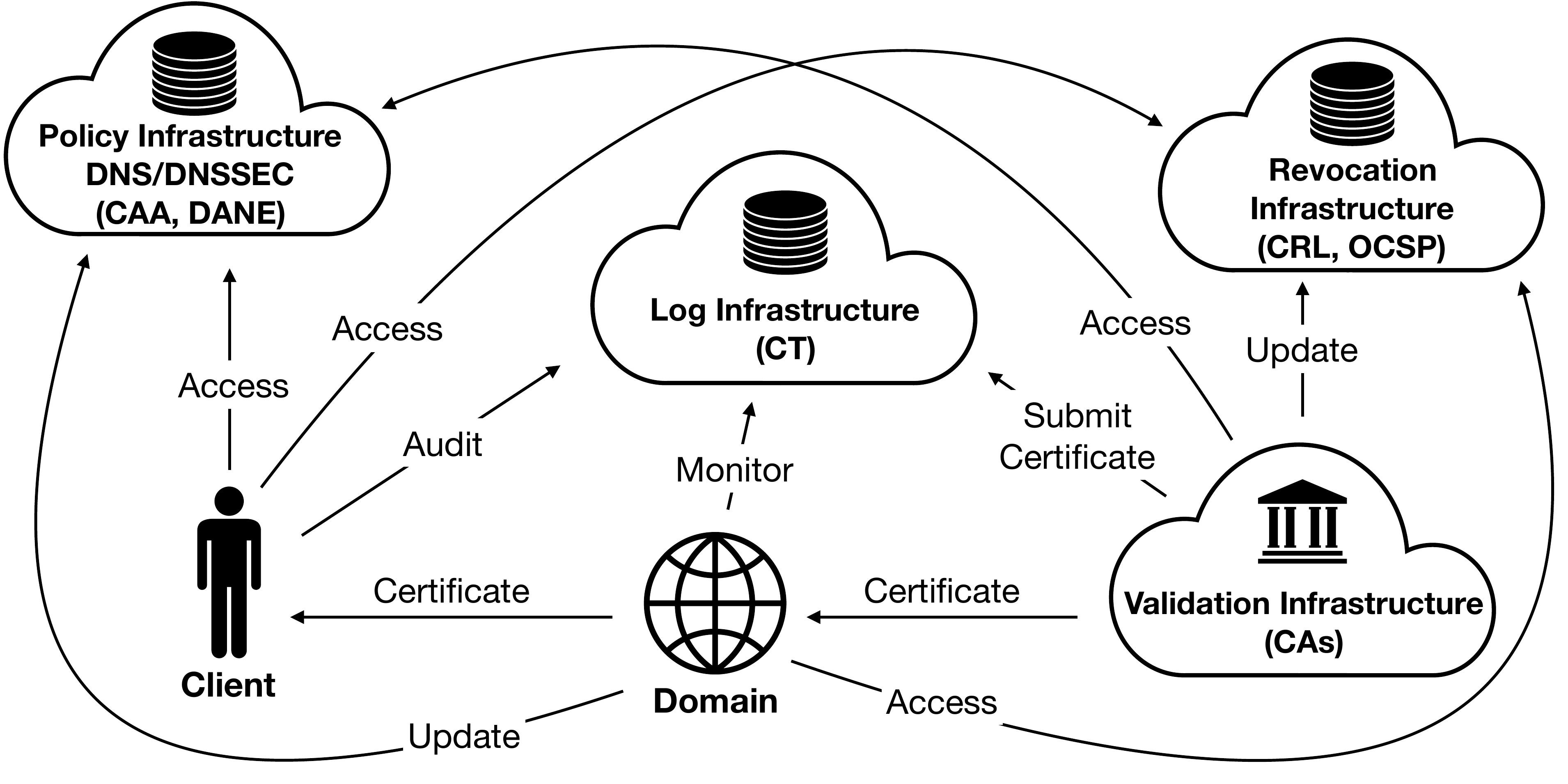}
    \caption{PKI infrastructures involved in the certificate validation.}
    \label{fig:current-pki}
\end{figure}

     A public-key validation is conducted only once per the certificate lifetime
     (i.e., prior its issuance) and certificates are ``static'', i.e., cannot be
     updated even if the claims they contain are outdated or incorrect.  For
     instance, a MitM adversary able to impersonate a domain even for a moment
     can obtain a certificate valid for several months.  Moreover, a certificate
     is  asserted only by a single (any trusted) CA what inherently limits its
     security.  In particular, an adversary compromising one CA can impersonate
     any domain on the Internet, as we have seen in the
     past~\cite{comodo-attacks,MozillaDigiNotarRemoval}. 
     Next, certificates are decoupled from the information about their validity,
     i.e., their content is insufficient to provide high guarantees to relying
     clients who validate them.  For instance, revocation information regarding
     a certificate is not part of the certificate and is stored and managed by a
     separate infrastructure.  Thus, to fully validate a certificate, a client
     has to obtain information from an external infrastructure. As a result, the
     certificate validation depends on availability and security of multiple not
     well-connected infrastructures.  To illustrate this complexity
     \autoref{fig:current-pki} depicts the currently deployed PKI
     infrastructures and their interactions.
     Certificates are inflexible in terms of introducing new
     functionalities and features, and their abilities are bound by their
     format which contains signed statements, assertions, and metadata produced
     at the time of their issuance. 
     For instance, current
     certificates cannot be authenticated by multiple CAs, effectively providing
     the same security level for all domains.  Another issue is a lack of domain
     expressiveness.  Domains cannot specify and enforce their security
     policies. In practice, it would be beneficial if domains could specify
     secure and flexible policies, like CAs authorized to issue their
     certificates or a number of CAs that certificates have to be asserted with.

\myparagraph{Design Rationale}
We argue that in order to
significantly improve the security of TLS PKI, a major redesign of digital
certificates and their ecosystem is necessary. 
%
     First, certificates should be \textit{self-contained},
        carrying all information needed to validate them, such that when a
        client receives a certificate, she should be able to validate it
        immediately without contacting any other parties or infrastructures.
     A framework should improve the \textit{security} of
        TLS connections. It should integrate the
        state-of-the-art security solutions and properties (like transparency)
        and
        certificates should offer
        stronger authentication of public keys (e.g., multiple validating
        CA) and it should be possible for domains to express and
        enforce their security policies (e.g., a list of trusted CAs).
        The framework should also be able to provide desired features
        impossible or difficult to provide now. One such example is
        \textit{key continuity}~\cite{key-cont} indicating how long
        a public key is being used by a domain.  Implementing such
        functionalities would imply that certificates are becoming rather
        ``dynamic'' as their internal state change over time.
     The framework should be \textit{efficient}.  For instance, many
        previous proposals for a revocation system has failed due to their
        inefficiencies~\cite{liu2015emc}. Lessons learned from these and other
        attempts are that: \textit{a)} clients should not incur
        excessive storage overheads, and \textit{b)} during TLS handshakes
        clients should not conduct any blocking connections to third parties.
     The framework should provide a viable and incremental
        \textit{deployment} path. In particular, it should be feasible to deploy
        the scheme without major changes to the ecosystem and existing
        protocols, or requiring new infrastructures. 

The main observation behind \name is that the blockchain and smart contract
technology allows us to rethink and redesign the concept of a digital
certificate, and to propose a system that meets the stated requirements.  As
depicted in \autoref{fig:current-pki}, today's certificate management and
validation is
underpinned by multiple distinct infrastructures with different goals and
operators; however, with similar requirements (like availability and
authentication).  We observe that functions of these infrastructures can
be easily implemented and potentially automated with the blockchain technology
and smart contracts.  First, blockchain platforms provide a good point of
anchoring trust as they are transparent, immutable, highly available, and
censorship resistant.  Second, smart contracts enable us to create objects
that can carry states updatable by a predefined code. In our case, we
can represent a certificate and its current validation state as such an object.
Consequently, the state can evolve, but only according to the domain's policy
and the rules encoded in the corresponding smart contract. For example, these
rules may require multiple authorized CAs to periodically conduct public-key
validations which then would be reported and checked by the code itself.  That
would result in self-contained and updatable certificates, where trust placed
in CAs is minimized as compliance of their actions is enforced and audited by
the code.  Lastly, the data models of blockchain
platforms enable light clients that do not have to store nor
process the entire blockchain but still can benefit from it.

\subsection{High-level Overview}
In \name, we propose to leverage smart contracts to create dynamic,
self-contained, extensible, and policy self-enforcing certificates.  The central
point of our design is to compose a digital certificate from its
\textit{validation} and \textit{presentation logic}.  The validation logic
maintains the current validation state of the certificate and its compliance
with the domain's policy, while the presentation logic presents this information
to clients.  Following this design, the following elements are central to
\name:

\begin{compactdesc}
    \item[\textbf{Policy contract}] is a smart contract that manages domains'
        security policies. Policies describe when domain's certificates are
        considered as valid. We assume that there is one global instance of this
        contract with a publicly known address.  Every domain can have a single
        policy at the time and if there is no policy for a domain the default
        policy is used.
    \item[\textbf{\name contract}] is a smart contract that implements, encodes,
        and enforces the public-key validation logic and policy compliance.
        The validation logic is expressed as a code (in the blockchain
        platform's programming language) with an associated storage. By
        executing the logic, the contract modifies its internal storage that
        reflects the validation state and its compliance with the domain policy
        (accessed from the policy contract).  \name contracts can implement
        arbitrary logic, 
        although we expect that only some, standardized,
        validation logic will be used in practice.
    \item[\textbf{\name certificate}] is an information that, in a secure and
        efficient way, presents TLS clients the current validation state of the
        contract.  Clients accept or reject TLS connections basing on \name
        certificates presented.
\end{compactdesc}
Throughout the paper, we present \name in a setting where \name certificates
carry information about historical public-key validations of a given domain
conducted by multiple CAs.  However, \name can be implemented with almost
arbitrary validation logic as it is only limited by the smart contract language
deployed.  We discuss other validation logic in
\autoref{pre:practical:policies}.  

\begin{figure}[b]
    \centering
    \includegraphics[width=0.95\columnwidth]{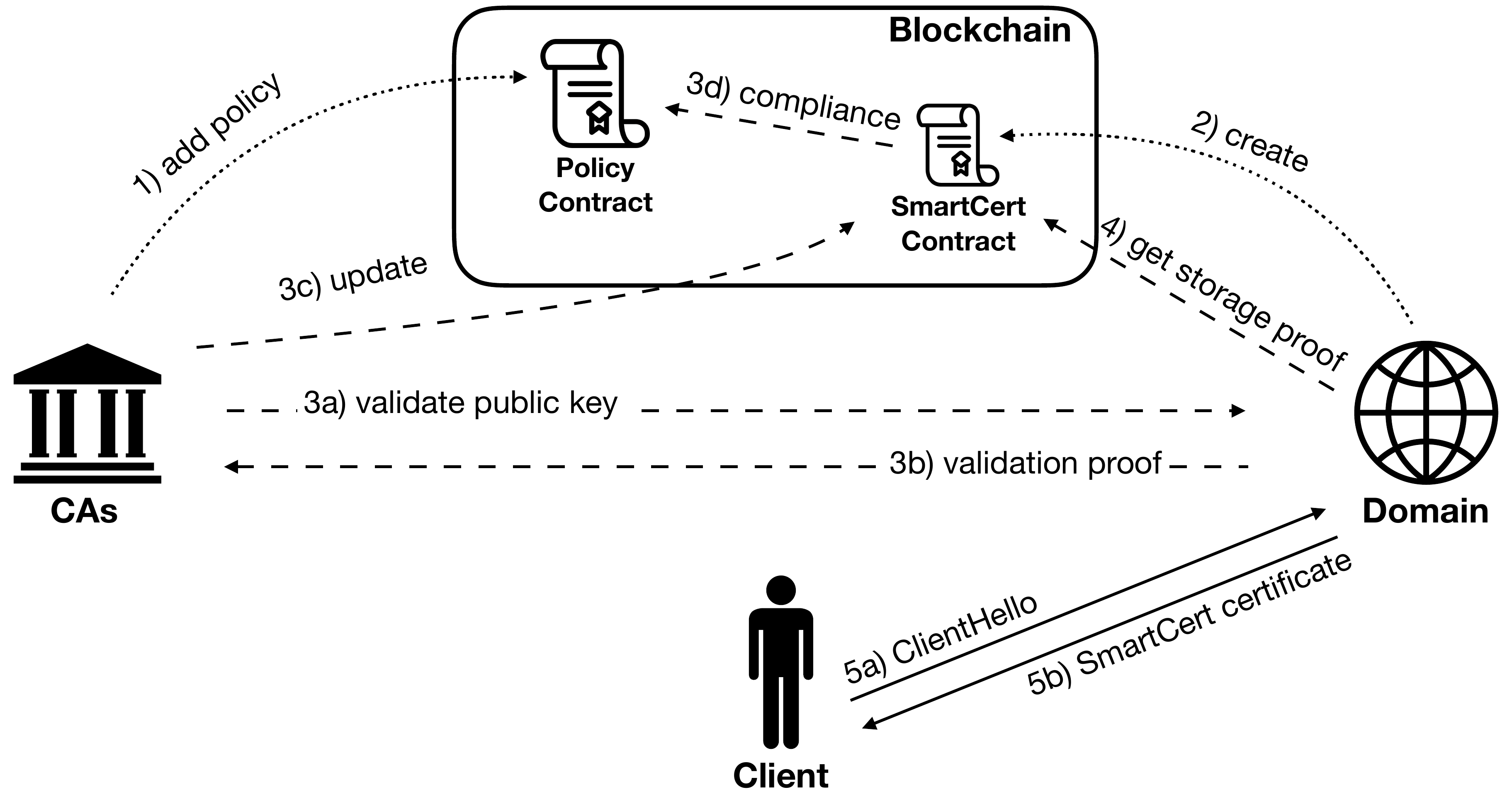}
    \caption{A high-level overview of the \name architecture. Dotted lines
    are for one-time operations, dashed lines for periodic actions, and
    solid lines for the TLS client-server communication.}
    \label{fig:overview}
\end{figure}

A high-level overview of \name is presented
in \autoref{fig:overview} and its steps are outlined below.

\begin{compactenum}
    \item In the first step, a domain creates its policy describing a set of
        conditions that its certificates will have to meet (e.g., what are
        authorized CAs or the maximum certificate lifetime).  To
        bootstrap the policy, the domain  contacts a number of CAs (at least
        one) with a request to sign the domain's policy. The CAs verify the
        domain's identity, sign the policy, and publish it at the
        policy contract. Policies can be updated subsequently; however, every
        domain can have only one ``active'' policy at the given time.
        The domain can skip this step and rely on the default policy.

    \item Then, the domain deploys a \name contract on the blockchain,
        created from a predefined (trusted) code template. It
        is initialized with the data like the domain name, the domain's public
        key, and the CAs that will be validating it. (We assume that the CAs are
        informed about the contract either by looking at the blockchain or
        out-of-band, e.g., via CA websites. In fact, the contract can be
        deployed by a CA instead of the domain.)
    
    \item Within every \textit{validation
        epoch} (or just \textit{epoch}), every CA listed in the \name contract
        is obligated to validate the domain's public key and update the
        contract:
        \begin{compactenum}
            \item a CA contacts the domain with a validation request,
            \item the domain returns a \textit{validation proof} which
                proves that at the time the domain possesses the private
                key corresponding to its public key from the contract,
            \item the CA updates the contract with the validation proof. The
                contract verifies it and updates its
                storage accordingly, reporting all occurred errors if any. 
            \item Finally, the \name contract checks if its validation state
                is compliant with the domain's policy and saves the outcome of
                this check as a \textit{validity status}.
        \end{compactenum}

    \item Periodically, the domain obtains from the blockchain network
        fresh MPT proofs  proving on the \name contract instance and its internal
        validation state.
        
    \item For every TLS connection, the domain sends to the client
        the \name certificate that consists of the MPT proofs accompanied with
        required metadata.  The client
        validates if \textit{a)} the corresponding \name contract was
        created correctly (i.e., has the correct code), \textit{b)} its validity
        status is correct, \textit{c)} and the MPT proofs are authentic and
        fresh.  Then, the client either accepts or rejects the TLS
        connection.  

        \name certificates carry historical information about public-key
        validations conducted, like the number of errors occurred or the last
        successful validation; thus, they can be combined with
        sophisticated policies or they can be used by domain operators to
        monitor validations.
\end{compactenum}

\section{\name Details}
\label{sec:details}

\subsection{Policy Creation and Update}
To facilitate domain expressiveness \name introduces a global policy contract
that contains a list of all trusted CAs (identified by their public-key hashes) and that allows domains to publish their
certificate validation policies.  By registering a policy, a domain specifies a
set of rules that domain \name certificates will have to be compliant with.
Whenever a domain wishes to register its policy it contacts  CAs (at
least one) that check the domain's identity, sign its policy, and publish it
with the global policy contract mapping domain names to their policies (see
\textit{newPolicy()} in \autoref{alg:policy}).  From that moment, everyone can
easily look up the domain's policy by calling \textit{getPolicy()} of the policy
contract.  If a domain does not register its policy, 
the default policy, with a relaxed security level (similar to today browser
policies), is used.  

\begin{algorithm}[t!]
    \caption{Policy Contract.}
	\label{alg:policy}
    \footnotesize
	\begin{inparadesc}
		\item[$\textnormal{\textit{dfltPolicy}}$:] the default
            policy\textbf{;}
		\item[$\textnormal{\textit{trustedCAs}}$:] the list of trusted CAs.
    \end{inparadesc}

    \SetKwProg{func}{function}{}{}
    \func{init()}{
        $\mathit{policies\leftarrow \{\}}$\;
    }

    \func{newPolicy(name, policy, sig[])}{
        $\mathit{p\leftarrow getPolicy(name)}$\; 

        /* policy update via policy key */\\
        \If{$\mathit{p\neq dfltPolicy}\textbf{\textit{ and
        }}\mathit{policy.type == UPDATE}$}{
            \textbf{assert} $\mathit{sender == p.KEYID}$;
            $\mathit{policies[name]\leftarrow policy}$\;
        }

        /* policy registration or replacement */\\
        \ElseIf{$\mathit{p==dfltPolicy}\textbf{\textit{ or
        }}\mathit{p.sigNo \leq len(sig)}$}{
            \textbf{assert} $\mathit{len(sig)>0}$; /* make sure at least one CA
            signs*/\\
            \For{$ca \in sig[]$}{
                \textbf{assert} $\mathit{ca \in trustedCAs}$\;
                \textbf{assert} $\mathit{SigVrfy_{ca}(name\|policy, sig[ca])}$\;
            }
            $\mathit{policy.sigNo\leftarrow len(sig)}$;
            $\;\mathit{policies[name]\leftarrow policy}$\;
        }
    }

    \func{getPolicy(name)}{
        \If{$\mathit{policies[name]}$}{
            \Return $\mathit{policies[name]}$;\quad/* return policy if registered */\\
            }
        \Return $\mathit{dfltPolicy}$;\quad/* return default */\\
    	}

\end{algorithm}

The main fields of a policy are:
\begin{inparadesc}
    \item[$\textnormal{\texttt{KEYID}}$] is a policy management public key
        identifier (unset in the default policy) that allows a domain to: update
        the policy, revoke any certificate claiming the domain name, and is
        mandatory for creation of any certificate claiming the domain name. As
        these actions are  infrequent,  the corresponding private key can be
        stored offline.
    \item[$\textnormal{\texttt{CAs}}$] is a subset of trusted CAs authorized to
        keep validating domain certificates (every trusted CA is authorized by
        default);
    \item[$\textnormal{\texttt{MAX\_LIFETIME}}$] is the maximum lifetime a
        domain's \name certificate can have (two years by default, as today); 
    \item[$\textnormal{\texttt{MAX\_ERR}}$] is the maximum number of validation
        errors per authorized CA to count this CA's validation as successful
        (unset by default);
    \item[$\textnormal{\texttt{MIN\_CAs}}$] is the minimum positive number of authorized
        CAs that validated a domain's \name certificate successfully (set to 1
        by default).
\end{inparadesc}

Each policy has other metadata associated, like its version, type (e.g.,
new or updated), and the number of distinct CAs which have signed it.
Security policies should be long-term and stable; however, domains from time to
time would need to modify them. Each policy includes its management key that can
be used for the policy update.  In the case of the key loss or compromise, the
policy can be replaced by a new policy signed by at least the same number of CAs
as the previous policy (see \textit{newPolicy()}).  At any given time, every
domain can have only a single ``active'' policy.  We discuss the policy contract
management and other possible policies in \autoref{pre:practical:policies}.

\subsection{\name Contract Creation}
To create a \name certificate, a domain has to create a \name contract
instance first.  Domains follow the same code template to instantiate \name
contracts, and we assume that clients are pre-loaded with the corresponding
code hash of that template (thanks to that, clients will be able to verify that
the received \name certificates were created using the authentic \name
contract code).  After the contract is appended to the blockchain it is
initialized (see \textit{init()} in \autoref{alg:contract}).  
First, the initialization procedure ensures that if the domain has its policy, the
\name contract is created with the private key corresponding to the
policy key (creating a binding between the policy and the contract).
Afterwards, the following contract fields are set:
\begin{inparadesc}
    \item[$\mathit{domainName}$] is the domain name for which the certificate is
    being issued;
    \item[$\mathit{pk}$] is the public key of the certificate;
    \item[$\mathit{created}$] denotes when the \name contract was created;
    \item[$\mathit{revoked}$] is the revocation flag;
    \item[$\mathit{valid}$] is the validity flag;
    \item[$\mathit{revs}$] is the set of CAs that revoked the certificate;
    \item[$\mathit{updated}$] denotes the time of the last \name contract
        update;
    \item[$\mathit{s[]}$] is a map of CAs authorized by the domain to keep updating the
        contract and their validation states with three following fields:
    \item[$\mathit{lastUpd}$] denotes the time of the last update of the given
        state;
    \item[$\mathit{lastErr}$] stands for the time of the last validation error
        (it can be either invalid validation proof sent to validate or no data
        sent, i.e.,  a skipped validation);
    \item[$\mathit{errNo}$] is the number of validation errors occurred so
        far.
\end{inparadesc}
After the initialization is completed, the \name contract instance is ready for
deployment.  

\begin{algorithm}[tb!]
    \caption{\name contract.}
	\label{alg:contract}
    \footnotesize
	\begin{inparadesc}
        \item[$\mathit{PC}$] is the global policy contract\textbf{;}\quad
		\item[$\textnormal{\textit{lastBHash}($i$)}$] returns the
            $i$-th last block hash\textbf{;}\quad
		\item[$\textnormal{\textit{lastBTime}($i$)}$] returns the
            $i$-th last block timestamp\textbf{;}\quad
		\item[$\textnormal{\textit{now}()}$] returns the timestamp of the
            current block\textbf{;}\quad
        \item[$\textnormal{\textit{cliRnd, srvRnd, params, $\sigma$}}$:]
            elements of the signed message \texttt{SKE} sent by the domain to
            the CA (see \autoref{sec:pre:tls}, \autoref{eq:call}, and
            \autoref{fig:update} for details)\textbf{;}\quad
        \item[$\mathit{s[]}$] is a map of CAs and their validation states.
    \end{inparadesc}
    \vspace{0.1cm}

    \SetKwProg{func}{function}{}{}
    \func{init(name, key, CAs)}{
        $\mathit{p\leftarrow PC.getPolicy(name)}$\; 
        \If{$\mathit{p\neq dfltPolicy}$}{
            \textbf{assert} $sender == p.KEYID$\;
        }
        $\mathit{created\leftarrow now()}$;
        $\;\mathit{domainName\leftarrow name}$;
        $\;\mathit{pk\leftarrow key}$\; 
        $\mathit{revoked\leftarrow False}$;
        $\;\mathit{revs\leftarrow\emptyset}$;
        $\;\mathit{valid\leftarrow True}$;
        $\;\mathit{updated\leftarrow 0}$\;

        \For{$ca \in CAs$}{
            \textbf{assert} $\mathit{ca \in p.CAs}$;
            $\mathit{s[ca].lastErr\leftarrow 0}$\;
            $\mathit{s[ca].errNo\leftarrow 0}$;
            $\mathit{s[ca].lastUpd\leftarrow now()}$\;
        }
    }

    \func{update(cliRnd, srvRnd, params, $\sigma$)}{
        $\mathit{ca \leftarrow sender}$;
        \textbf{assert} $\mathit{ca \in s[]}\textbf{\textit{ and }}\mathit{valid}$\;

        /* check for skipped validations of CAs*/\\
        \For{$tmp \in s[]$}{
                $\mathit{missed\leftarrow \lfloor{\frac{now() -
                s[tmp].lastUpd}{epoch}}\rfloor}$\;
                \If{$\mathit{missed\geq 1}$}{
                    $\mathit{s[tmp].lastErr \leftarrow now() - epoch}$\;
                    $\mathit{s[tmp].errNo \leftarrow s[tmp].errNo + missed}$\;
                }
        }

        /* ensure that \textit{cliRnd} is fresh */\\
        $\mathit{epochStart = now() - (now() \% epoch)}$;
        $\;\mathit{fresh\leftarrow False}$\;
        \For{$i \in \{1,2,3,...\}$}{
            /* check if the $i$-th last block is in the epoch */\\
            \If{$\mathit{lastBTime(i)<epochStart}$}{
                \textbf{break}\;
    		}
            /* check if the $i$-th last block hash matches */\\
            \If{$\mathit{cliRnd==ca\|lastBHash(i)}$}{
                $\mathit{fresh\leftarrow True}$;
                \textbf{break}\;
    		}
    	}

        /* check the freshnes and verify the signature */\\
        \If{\textbf{!}fresh \textbf{or}
            \textbf{!}$\mathit{SigVrfy_{pk}(cliRnd\|srvRnd\|params,\sigma)}$}{
            $\mathit{s[ca].lastErr\leftarrow now()}$;
            $\;\mathit{s[ca].errNo}$++\;
    	}
        $\mathit{updated\leftarrow now()}$;
        $\mathit{valid\leftarrow isCompliant()}$\;
    }

    \func{revoke()}{
        $\mathit{p\leftarrow PC.getPolicy(name)}$; 
        $\;\mathit{r\leftarrow 0}$\; 
        \If{$\mathit{sender \in p.CAs}$}{
            $\mathit{revs.add(sender)}$;
            $\;\mathit{r\leftarrow len(p.CAs)-len(revs)}$\; 
        }
        \If{$\mathit{sender == p.KEYID}\textbf{\textit{ or
        }}\mathit{r\geq p.MIN\_CAs}$}
        {
        $\mathit{updated\leftarrow now()}$;
        $\;\mathit{revoked\leftarrow True}$;
        $\;\mathit{valid\leftarrow False}$\;
    }
    }

    \func{isCompliant()}{
        $\mathit{i\leftarrow 0}$;
        $\;\mathit{p\leftarrow PC.getPolicy(name)}$\; 
        \For{$ca \in s[]$}{
            \If{$\mathit{(ca \in p.CAs)}\textbf{ and }\mathit{(s[ca].errNo
            \leq p.MAX\_ERR)}$\\
           \hspace{0.33cm}\textbf{and} $\mathit{(now() - created \leq
           p.MAX\_LIFETIME)}$
            }{
            $\mathit{i}$++\;
            }
        }
        \Return $\mathit{i \geq p.MIN\_CAs}$\;
    }
\end{algorithm}

\subsection{Public-Key Validation and Contract Update}
\label{sec:details:valid}
In our scenario, \name certificates carry information about historical
validations of their public keys associated.  To realize it, every
\textit{epoch} (a system parameter discussed in \autoref{pre:practical:params}),
each authorized CA conducts the public-key validation with the domain and submits the
obtained \textit{validation proof} to the corresponding \name contract to update
it.  The goal of this procedure is to ``convince'' the \name contract that the
domain still possesses the key pair, such that the contract can verify it and
update its storage to reflect the validation state accordingly. Such 
a proof should be:
\begin{inparaenum}[\itshape a)]
    \item short, so a smart contract can process it efficiently
    \item \textit{authentic},  proving that the private key,
        corresponding to the public key specified in the \name contract, is
        being used,
    \item \textit{fresh}, proving that the information was produced
        recently.
\end{inparaenum}

The authenticity is necessary as it should be infeasible to produce fake
validation proofs that a given key in being used. Moreover, a \name contract
must be able to verify that proofs are authenticated by the
private key associated with the public key that the contract was initialized
with. On the other hand, the freshness is critical, as otherwise a malicious CA
could just produce multiple validation proofs at once and keep submitting them
later to the \name contract over some extended time period. Therefore, a \name
contract must be able to verify that the validation proof is recent.  However,
these two requirements are challenging to meet in the TLS setting.

\myparagraph{Authenticity}
The TLS protocol was not designed to provide non-repudiation for
application-level data. Therefore, it is impossible to prove to third parties
that a server produced given data.  However, such proof is necessary for our
setting, as \name contracts have to be able to verify that a given private key is
being used.
One possible approach to tackle this problem is to design and deploy a new
protocol or an extension that would allow servers to sign 
date~\cite{barnes2017automatic,ritzdorftls}.  However, that may be undesired as introducing a
new dedicated protocol have side effects, like larger trusted computing base and
a long standardization and adoption process.
Instead of introducing a new protocol, we make an
observation that although TLS does not provide non-repudiation for application
data, the TLS handshake messages can be used to achieve this goal.  More
specifically, as described in \autoref{sec:pre:tls}, during the TLS handshake
with a negotiated ephemeral DH key exchange method, the server signs client's
and server's random values (\CHrnd and \SHrnd, respectively) together with the
DH parameters. The signature is placed in the ServerKeyExchange message
(\texttt{SKE}) sent to the client; therefore, it can be used as a proof that the
server's private key is being used.  CAs can obtain such proofs by simply
running TLS handshakes with targeted domains.

\myparagraph{Freshness}
The message authenticity is insufficient for a complete validation proof. A CA
could conduct multiple handshakes at once and keep using them later on as
``current'' validation proofs.  Thus, the second challenge is to provide
freshness for that information.  
Interestingly, the client's and server's random values as specified in the TLS
specification~\cite{rfc5246}, should contain a GMT timestamp encoded in the
first four bytes of these fields (see \autoref{fig:tls_handshake}).  Therefore,
the ServerKeyExchange message containing the server's random value contains also
the server's timestamp.  Unfortunately, we cannot rely on this field, as
only 37\% of TLS servers put accurate
timestamps in their TLS handshakes~\cite{szalachowski2018blockchain}. Moreover, the TLS protocol
does not require protocol parties' clocks to be synchronized~\cite{rfc5246}
and some implementations randomize timestamp fields as a
prevention from fingerprinting~\cite{remove_gmt}. 

One approach to solving this issue is to use external time sources to timestamp
TLS handshakes but that would require the introduction of new trusted parties
and reliance
on their availability.  To overcome that problem, we make an observation
that the blockchain platform itself acts as a source of time.  
Blockchain blocks are ordered (linked by a hash function) and
contain their creation timestamps (see \autoref{fig:state}).
Therefore, a CA could set its ClientHello's random field (\CHrnd) to the
current block hash and conduct a TLS handshake.  Then when this value is signed
by the server (within the \texttt{SKE} message), it implies that the handshake
happened after the block was mined.  Therefore, the \name contract will be able to
verify that the validation proof is fresh, checking whether the signed block
hash is recent (smart contract platforms, like Ethereum, natively
support accessing previous block hashes). In that way, the validation will
be automated and without introducing any additional trusted parties or protocol
changes.

\begin{figure}[t]
    \centering
    \includegraphics[width=\columnwidth]{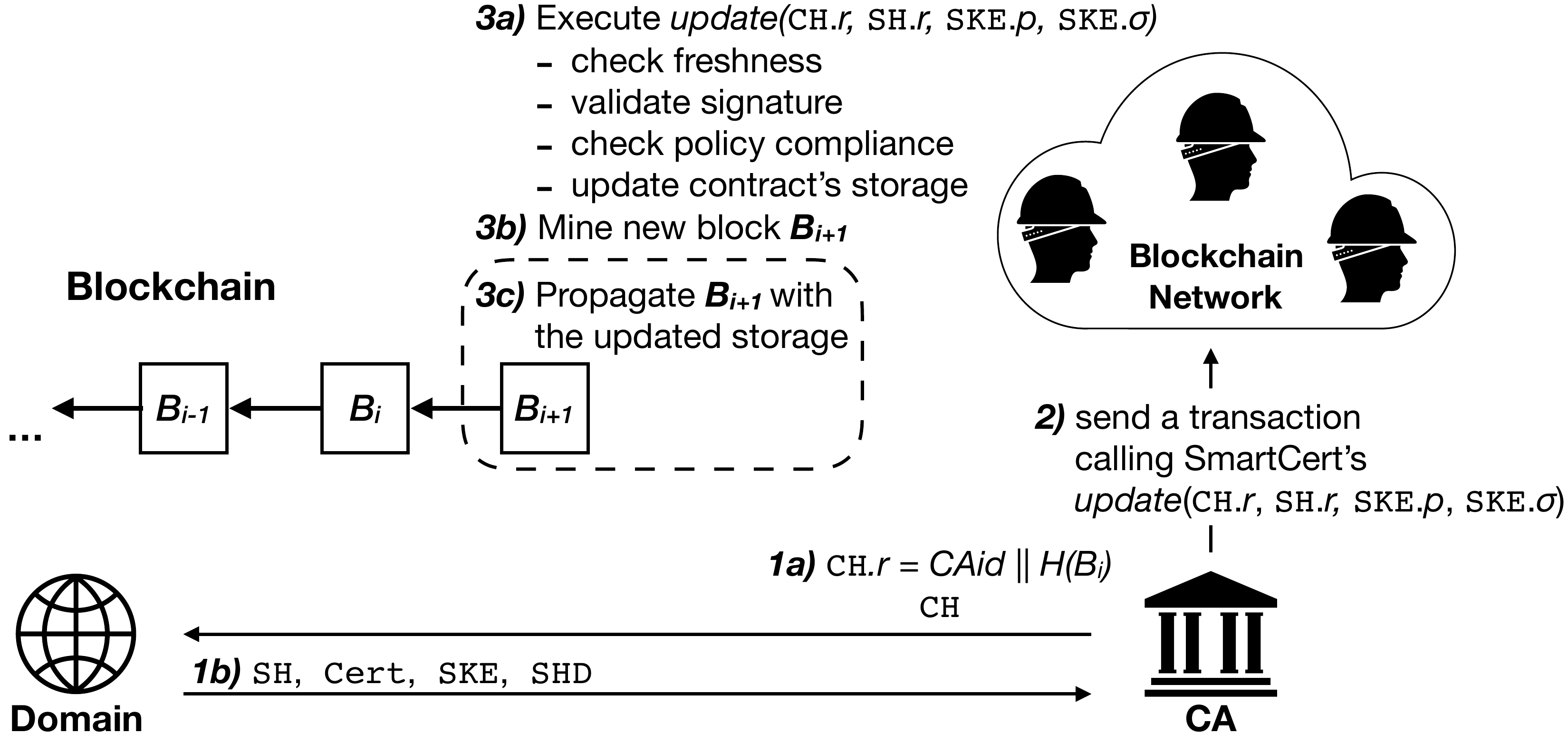}
    \caption{Details of the \name contract update process.}
    \label{fig:update}
\end{figure}

\myparagraph{Public-Key Validation and Contract Update}
The public-key validation is initiated by a CA once per epoch.  The entire
process is depicted in \autoref{fig:update} and \autoref{alg:contract} (see
the \textit{update()} method), and we discuss it in detail below.

\begin{compactenum}
    \item The CA and domain execute a TLS handshake:
    \begin{compactenum}
        \item The CA gets the latest block ($B_i$) hash, creates a
            hello message \texttt{CH}, prepends the hash with its
            blockchain identifier
            ($\mathit{CAid}$) and sets it as the random field of the message:
            \begin{equation}
                \label{eq:block_hash}
                \CHrnd = \mathit{CAid}\|H(B_i).
            \end{equation}
            The CA identifier $\mathit{CAid}$ is added to ensure that any other
            CA cannot replay the validation proof.
            The message \texttt{CH} is sent to the domain.
        \item The domain sends back the messages
            \texttt{SH},\texttt{Cert},\texttt{SKE}, and \texttt{SHD}. The
            ServerKeyExchange message \texttt{SKE} contains the signature over the
            random fields and the key exchange parameters (see
            \autoref{fig:tls_handshake} and \autoref{sec:pre:tls}).
            This signature together with the fields it protects constitute a
            validation proof.
    \end{compactenum}

\item After receiving the proof, the CA prepares and disseminates a
    transaction to trigger the \textit{update()} method of the \name contract
    (see \autoref{alg:contract}), that will be executed with the elements
    of the validation proof as arguments: 
        \begin{equation}
            \label{eq:call}
            \mathit{update(\CHrnd,\SHrnd,\texttt{SKE}\textit{.p},
            \texttt{SKE}.\sigma)},
        \end{equation}
        where \CHrnd is set to the block hash set by the CA. 

    \item Network participants (i.e., miners) constantly collect new transactions and try to mine a new block.
    \begin{compactenum}
        \item Miners validate the transaction by executing
            \textit{update()}. 
            The method checks if
            the submitter is an authorized CA,  if the certificate is
            still valid, and executes the following verification
            logic.
            \begin{compactenum}
                \item Check if any authorized CA skipped a validation and
                    every missed validation count as an error.
                \item Check whether the message is fresh
                    by comparing \CHrnd with the previous block hash(es).  Due
                    to the network propagation
                    time~\cite{weber2017availability}, multiple older block
                    hashes can be checked (see \autoref{pre:practical:params}
                    for more details); however, all these
                    blocks must belong to the current epoch (to avoid replayed
                    validation proofs).  If the matching block hash is found,
                    it implies that the TLS handshake was conducted after
                    the block was mined (as the hash is within the \CHrnd
                    field); thus the validation proof is fresh.
                \item Verify the authenticity of the handshake's
                    \texttt{SKE} message, i.e., ensure that it was signed by the domain's private key
                    corresponding to the public key with which the contract was
                    initialized.
                \item Update the contract's internal storage indicating
                    that it just got updated (i.e., the last validation
                    timestamp is updated). If any of the above checks fails, the
                    counter of errors is increased and the last error time is
                    updated.  
                \item Determine the validity status
                    by checking if the validation state is compliant with
                    the domain's policy (see \textit{isCompliant()}). In this
                    check, the contract gets the domain's policy via the policy
                    contract and checks if \textit{a)} the CAs updating the
                    contract are authorized, \textit{b)} their
                    validation states are correct, \textit{c)} the certificate
                    lifetime is not exceeded, \textit{d)} and the number of
                    correct validations is at least as specified in the policy.
                    If the policy compliance check fails the
                    certificate will be marked and treated as invalid (i.e.,
                    \textit{valid=False}).
            \end{compactenum}
            At the end of this process, the \name contract has validated the
            authenticity and freshness of the handshake message and updated its
            internal storage accordingly to the domain's current policy. 

        \item Miners mine block $B_{i+1}$ with validated transactions.
        \item When the new block $B_{i+1}$ is mined and propagated it
            becomes part of the blockchain. The block contains the
            updated state of the \name contract.
    \end{compactenum}
\end{compactenum}
The update procedure requires an interaction with a smart-contract platform.
As we show in \autoref{sec:impl:eval} even with the existing platforms the
execution cost is acceptable.

\myparagraph{Revocation}
Like in the TLS PKI, certificates can be revoked when they are issued by a
mistake, their associated keys are lost/stolen or not used anymore, or their
owners just wish to revoke them.  A \name certificate can be revoked by a quorum of
its authorized CAs or the domain owner.  In order to do so, the \textit{revoke()}
method is triggered, which verifies whether the calling entity is authorized and
upon a successful revocation sets the \textit{revoked} flag to \textit{True} and
the \textit{valid} flag to \textit{False}. With that change, the \name
certificate is revoked and immutable, as the \textit{update()} method checks the
validity status prior to executing the rest of its body.

\subsection{\name Certificate and its Validation}
\label{sec:details:cert_valid}
\name does not change the TLS protocol and it follows the standard TLS
connection establishment as described in \autoref{sec:pre:tls}. The only change is
that instead of sending an X.509v3 certificate, the server sends a \name
certificate, which consists of the following elements:
\begin{compactdesc}
    \item[$\mathit{addr}$] is the blockchain address of the \name contract,
    \item[$\mathit{st}$] is the storage of the \name contract that encodes the
        domain name, its public key, and its current validity status (i.e.,
        whether the certificate is valid or not),
    \item[$\pi_s$] is an \textit{MPT storage proof} proving that the
        given contract instance's storage is part of the blockchain.
    \item[$\pi_c$] is an \textit{MPT code proof} proving on the code
        of the \name contract (without this proof a
        malicious CA could deploy a contract with a different code).
\end{compactdesc}
To prevent replay attacks, the storage and the proofs have to be fresh, so
the domain periodically, e.g., every several hours, fetches new proofs
from the blockchain and serves them in every subsequent TLS handshake.
We emphasize that this setup is privacy preserving as clients do not contact
the network for
fresh proofs (revealing domains they connect to). 
The \name certificate validation is described in \autoref{alg:validation}.  In
the first step,
the client verifies if:
\begin{inparaenum}[\itshape a)]
    \item the MPT proofs provided are consistent and authentic, i.e., they
        correspond to the same smart contract address $\mathit{addr}$ and they belong to the
        blockchain whose block headers the client stores
        (TLS clients are also blockchain light client), 
    \item the code proof $\pi_c$ proves on the correct smart
        contract code (CAs have to follow the same code template and the
        clients are pre-loaded with the hash of this template
        \textit{codeHash}), 
    \item the storage proof $\pi_s$ proves that the passed storage $st$
        belongs to the contract.
\end{inparaenum}

\begin{algorithm}[t!]
    \caption{\name certificate validation.}
	\label{alg:validation}
    \footnotesize
	\begin{inparadesc}
        \item[$\textnormal{\textit{codeHash}}$] is the hash
            of the trusted \name contract code stored by clients\textbf{;}\quad
      \item[$\textnormal{\textit{maxStale}}$]  denotes the maximum tolerable
          difference between the current time and the time of the last update
          of the presented certificate (see \autoref{pre:practical:params}),
          $\mathit{maxStale>epoch}$.
    \end{inparadesc}
    \vspace{0.1cm}

    \SetKwProg{func}{function}{}{}
    \func{verify(name, addr, st, $\pi_s, \pi_c$)}{

        /* check proofs authenticity and consistency */\\
        \If{\textbf{!}$\mathit{AuthConsistentProofs(addr,\pi_c,\pi_s)}$ \textbf{or}\\
           \hspace{0.3cm}\textbf{!}$\mathit{codeProofVrfy(addr, codeHash,\pi_c)}$  \textbf{or}\\
           \hspace{0.3cm}\textbf{!}$\mathit{storageProofVrfy(addr, st,\pi_s)}$
        }{
            \Return FAIL\;
    	}

        /* check the name, validity, and freshness */\\
        \If{$\mathit{st.domainName\neq name}$ \textbf{or !}$\mathit{st.valid}$ \textbf{or}\\
            \hspace{0.3cm}$\mathit{time()-st.updated > maxStale}$
            }{
            \Return FAIL\;
        }

        \Return OK;\quad /* continue the handshake with $\mathit{st.pk}$ */
    }
\end{algorithm}

When the client knows that the storage corresponds to an authentic \name
contract instance, the client verifies whether \textit{a)} the \name
contract is issued for the contacted domain, \textit{b)} the certificate is
valid (i.e., compliant with the policy and non-revoked), \textit{c)} the storage is fresh
enough i.e., was updated no longer than $\mathit{maxStale}$ seconds ago.
The validity of the \name certificate implies its compliance with the domain's
policy (note, that the correct code template guarantees that the contract keeps
checking its compliance with the correct policy).  In our example, the \name
contract self-enforces the policy by checking if the number of public-key
validation errors does not exceed the maximum number allowed and if
validations were successfully conducted by the minimum number of
authorized CAs as specified by the domain.  Interestingly, with changing
policies and \name contracts, TLS clients stay almost unchanged, as they only
check the validity of the certificates while the compliance checks and
validations are conducted by \name contracts themselves.
After the certificate is successfully validated, the client continues the TLS
handshake with the domain's public key
specified in the \name contract storage (i.e., $\mathit{st.pk}$).

\section{Security Analysis}
\label{sec:analysis}
\name provides multiple security benefits over standard certificates.  First of
all, \name provides a reliable and easy, yet powerful, policy framework
supporting domain expressiveness.  Domains can publish their policies specifying
conditions under which \name certificates are considered valid. Policies can
be authenticated by multiple CAs (at least one); thus, \name is not a
weakest-link system anymore. 
Furthermore, if a domain's policy is published it guarantees that
\textit{a)} no \name contract (and no \name certificate, consequently) can be
created without the domain's consent, and \textit{b)} all \name contracts
claiming the domain's name will self-enforce the policy.  For the former, if
there exists a policy for a domain name,  a \name contract with that name
specified can be created  only when it is authenticated by the domain's policy
key (see \textit{init()} in \autoref{alg:contract}).  For the latter, \name
contracts with every update check their compliance with the policy; thus, any
non-compliant contract will invalidate itself with the first update and
consequently will be rejected by clients (note, that clients check if the \name
contract code is correct to ensure that the validation rules are respected).
An adversary can present such an invalidated \name certificate to clients, but
only for \textit{maxStale} seconds from its latest update (then clients will
reject it -- see \autoref{sec:details:cert_valid}).

Beside authenticating policies by multiple authorities, valid \name contracts
have to be successfully authenticated by a minimum \texttt{MIN\_CAs} CAs from the
\texttt{CAs} set, as specified by the domain.  Therefore, it increases the
security as certificates are being validated by multiple authorized CAs and it
limits adversary's capabilities as an adversary compromising a CA can only
update certificates that have authorized this particular CA.  An adversary able
to compromise multiple ($m$) CAs can try to impersonate a domain by updating its
policy. Such an attack will be successful if $m$ is at least equal the number of
CAs that signed the current legitimate policy. However, as assumed the domain
can obtain a new policy signed by $n>m$ CAs, and submit it to invalidate the
malicious policy.   As policy keys are used infrequently (only for policy
updates, \name contract creations, and revocations) they can be stored offline
increasing the security.  However, in the case of key loss or compromise, the
same policy update procedure is executed. Besides, all policy changes are
publicly visible so the domain should be able to quickly mitigate these threats.

If a domain deploys \name certificates without a policy specified, the default
policy is used. It authorizes any trusted CAs to validate domain certificates,
so
 an adversary able to compromise a single CA can impersonate the domain
without a policy.  Such an attack would be detectable easily, as the malicious
contract would be published on the blockchain. In such a case, the
domain can quickly bootstrap its policy and revoke a certificate claiming its
domain name via the policy key.  The adversary would not get a
successful validity status for the revoked certificate (the \name contract
would self-invalidate with the successful revocation), but she would be
able to use the old proofs for at most $\mathit{maxStale}$ seconds.

\name contracts encode their current validation states; therefore, it is easy to
get a better view of public keys of contacted domains.  More specifically, in
the presented setting, a \name contract encodes the key continuity measure and
checks its compliance with the domain's policy, and then reflected the
compliance in the \name certificate.  CAs are obligated to conduct public-key
validations periodically, and this procedure is enforced and monitored by the
smart contract's code. Such a design allows to provide the functionality of
notary systems detecting and preventing various MitM attacks where e.g., an
adversary impersonates the website presenting a different certificate. If a CA
is presented with the malicious key, it will be visible as it will be reflected in the
\name contract. Moreover, a CAs without breaking the
blockchain platform cannot prefetch validation proofs as they require recent
pseudorandom block hashes.

\name contracts are designed such that only authorized CAs can update them and
such updates are validated by the code and individual domain policies.
Therefore, \name minimizes trust placed in CAs.  If a malicious or unavailable
CA does not submit a validation proof within an epoch or submits an invalid
validation proof, the \name contract will detect it and report that fact in its
internal storage.  Similarly, a replayed or outdated validation proofs will not
be accepted by the \name contract as valid and an error will be recorded as
well.  Consequently, \name provides an important property that if a
domain is under a MitM attack during an epoch, it will be reflected in the
corresponding \name contract and issuing CA cannot refuse or hide this fact.
This design is beneficial not only for TLS clients but also for domains who can
detect attacks much easier just by looking at their \name contracts (see
\autoref{pre:practical:policies}).

\name benefits from the underlying blockchain platform in other ways.  An
immediate benefit is a high availability, which turned out to be an
\textit{Achilles heel} of almost all security
infrastructures. 
\name certificates are
created and distributed via the blockchain platform, and they contain all
information required to validate them (like a validation state and a revocation
status).  Domains obtain fresh MPT proofs just by interacting with the
blockchain network, which is distributed, available, and censorship resilient.
CAs need to be able to only send transactions to the blockchain network, but
they  do not need to invest in any highly-available front-end servers to serve
TLS clients.  It increases the security of the system significantly, as an
adversary trying to launch a Denial-of-Service attack on a CA or wanting to
censor CA transactions needs to undermine the properties of the blockchain
platform what requires a (significant) control over the blockchain network.
Another important benefit of basing \name upon the blockchain platform is
transparency.  All policies, certificates, and other CA actions are publicly
visible.  Therefore, \name allows to detect CA mistakes or tailored attacks
where a fraudulent certificate is created by a malicious or colluding CA for a
one-time targeted attack.   In \autoref{pre:practical:policies} we discuss how
smart-contract platforms facilitate such a monitoring.  In addition, parties in
\name cannot equivocate, e.g., by registered two policies for the same domain,
as long as they cannot undermine the properties of the blockchain platform.
Finally, \name is privacy-friendly, since \name certificates are self-contained and sent
to clients by servers during TLS handshakes. Thus, clients do not need to
contact any other parties (like CAs) or services to fully validate \name
certificates.


\section{Implementation and Evaluation}
\label{sec:implementation}
%
\subsection{Implementation}
We implemented all \name elements, namely a \name-supported
client, server, CA, and smart contracts.  Although \name can be implemented with
more performant permissioned blockchains (see \autoref{sec:limitations}), we
present it in a more challenging open setting; thus we selected Ethereum as the
blockchain platform and implemented the contracts in Solidity.  For
in-contract signature verification, we used and extended the
\texttt{SolRsaVerify} package, which provides a verification code for
RSA signatures used in ServerKeyExchange messages (i.e., RSA-PKCSv1.5-SHA256~\cite{rfc3447}).  We implemented RSA signatures since they are most
common as a TLS authentication method~\cite{huang2014experimental}.
The CA is implemented in Python and it consists of \textit{a)} a blockchain-communication module
to interact with the Ethereum platform and contracts, and \textit{b)} a
modified TLS stack that obtains validation proofs from TLS handshakes.
The former module is realized with the
\texttt{Web3py} package that allows CAs to get a recent block hash (used in
public-key validations --- see \autoref{sec:details:valid} and
\autoref{alg:contract}) and send transactions to policy and \name contracts. For
modifying
the TLS stack we used the \texttt{tlslite-ng} library.  With
our changes, prior a TLS handshake, a TLS client can pass an arbitrary
string as its hello's random value (\CHrnd), and then get it signed
together with the fields of the ServerKeyExchange message (i.e., \SHrnd,
\texttt{SKE}\textit{.p}, and $\texttt{SKE}.\sigma$).  
%
We implemented a \name server with NGINX 1.15.2. Our server differs from a
standard TLS server by two following elements: \textit{a)} it periodically
obtains fresh MPT proofs, and \textit{b)} it transfers its \name certificate to
the TLS client.  The first functionality is implemented with a help of the
\texttt{eth-proof} package, while the second one is realized by employing the
OCSP stapling extension~\cite{rfc6961} allowing a server to piggyback a
\name certificate during a TLS handshake.  Moreover, such an implementation is
backward-compatible as \name-unsupported clients can ignore the extension (see
\autoref{pre:practical:deploy}).
\name client, realized mainly in Python,  implements the TLS connection
establishment, parsing of the OCSP stapling extension, and the \name
certificate validation logic (see \autoref{alg:validation}), including the
verification of MTP proofs (via the \texttt{eth-proof} package).

\subsection{Evaluation}
\label{sec:impl:eval}

\myparagraph{Smart Contracts}
  Deploying and interacting with
smart contracts over Ethereum is associated with a gas cost. Each operation
costs a pre-defined amount of gas, and operations can differ significantly in
their gas consumption.  To estimate the cost in US dollars we used the following
data from \url{https://ethgasstation.info/} obtained on 20 Mar 2020: we used the
\textit{standard gas price} equal 1.8 Gwei/gas (1 Gwei = $10^{-9}$ ether) with the
ether price equal \$150/ether.  
We deployed our smart contracts on an Ethereum testnet.
First, we measured the cost incurred by registering a policy with the policy
contract (see \textit{newPolicy()} in \autoref{alg:policy}).  The overall cost
is around \$0.23 and it is
mainly due to the transaction size and the storage required for the policy
fields. 
%
Next, we investigated how expensive it is to deploy \name certificates.  In our
experiment, the contract was deployed for \url{facebook.com}, with
which a CA kept conducting TLS handshakes periodically (as presented in
\autoref{sec:details:valid}) and kept submitting obtained validation
proofs to update the \name contract.  \url{Facebook.com} signs ServerKeyExchange
messages with RSA-2048 and the results of this experiment are presented in
\autoref{tab:cost}.  As shown in the table, creating and initializing the contract is more expensive than updating it, and according to
our estimation, it costs around \$0.36 (it is a one-time cost
per \name contract).  To update a \name contract by a validation proof, a CA has
to pay about \$0.045 in that scenario.  As depicted, the RSA verification is the
\textit{update()}'s dominant operation, consuming almost 56\% of all gas
required.
\begin{table}[t]
\begin{center}
\footnotesize
\caption{Cost analysis of the \name contract operations.}
\label{tab:cost}
\begin{tabular}{|c||r|r|r|r|}
    \cline{3-5}
    \multicolumn{1}{c}{} &
    \textbf{create}&\multicolumn{3}{|c|}{\textbf{\textit{update()}}} \\
    \cline{3-5}
    \multicolumn{1}{c}{} &
    \textit{\textbf{+ init()}} & \textbf{Total } &
    $\textbf{\textit{SigVrfy()}}$ & \textbf{Misc.\ \ }\\
    \hline
    \hline
    Gas&1343799&165885&92623&73261\\
    ETH&0.00289&0.00036&0.0002&0.00016\\
    USD&\$0.361&\$0.045&\$0.025&\$0.020\\
    \%&100\%&100\%&55.8\%&44.2\%\\
    \hline
\end{tabular}
\end{center}
\end{table}
%
%
The cost of maintaining a \name certificate is
mainly determined by the epoch parameter (see discussion in
\autoref{pre:practical:params}). In \autoref{tab:month} we present a monthly
cost depending on epoch values assuming a single CA (the cost is
linear to the number of CAs).  As shown, the cost is inversely proportional to
epoch, e.g., the monthly cost for a \name certificate validated every
day is \$1.26.

\begin{table}[t]
\begin{center}
    \small
    \caption{A monthly deployment cost depending on epoch values.}
    \label{tab:month}
    \begin{tabular}{c|rrrrr}
        \textbf{Epoch}    & 6h  &   12h   &   24h  &  48h &  72h  \\ \hline
        \textbf{Cost}     & \$5.05  &\$2.53 & \$1.26 &\$0.63 & \$0.42 \\
    \end{tabular}
    \end{center}
\end{table}

Ethereum, like other blockchain platforms, is under heavy development.
Therefore, we envision that with scaling solutions and more mature
implementations (e.g., natively-supported RSA operations), the gas consumption
presented may be much lower in future, consequently lowering the overall cost of
\name deployment.

\myparagraph{CAs} 
The main overhead on the CA's side is to conduct periodic validations which is
equivalent to running TLS handshakes (see
\autoref{sec:details:valid}).  To estimate the throughput, we
selected 10000 random hosts supporting TLS and we
conducted periodic TLS handshakes with them.  We used a machine equipped with
Intel Core i7 (3.5GHz), 16GB RAM, and 100Mbps wired Internet connection.
This single machine running connections in parallel could perform
around 175 TLS handshakes per second on average, i.e., 0.63M handshakes per
hour (the time is dominated by the network latency). Thus, assuming six hours
epoch,
only three machines could keep validating over ten million
public keys, which is similar to the number of active TLS certificates
currently issued by even large CAs~\cite{vandersloot2016towards,ca_usage}.

\myparagraph{Transmission Overhead}
\name certificate size is the only transmission overhead introduced to a
TLS connection. To estimate it we used a \name
certificate derived from the \name contract deployed previously.
The main contributors to the \name certificate size are MPT storage and code 
proofs ($\pi_{s}$ and $\pi_{c}$, respectively).  MPT proofs in Ethereum are
generated for 32B-long words; therefore, a domain's public key (usually,
256B long) requires multiple MPT proofs.  However, these
proofs are highly redundant, and in our scenario, the \name certificate needs
4.11KB to be encoded.  As reported~\cite{VaSz18}, an average size of today's
X.509v3 certificate chain is around 4.75KB; thus, the size of \name
certificates is comparable with today TLS certificates. 

\myparagraph{TLS Clients}
In \name, TLS clients are also blockchain light clients
obtaining and verifying new block
headers (without actual blocks and transactions) from the blockchain network.
In Ethereum, block headers are about 508B long; thus, at the time
of writing, the Ethereum main network with 6.75M blocks requires
from \textit{fully light clients}~\cite{light_cli} the storage overhead of about
3.43GB.  However, storing all previous header is not necessary for validating
\name certificates.  TLS clients do not accept \name certificates older than the
\textit{maxStale} parameter. So, if a client rejects stale certificates with
storage proofs older than three days, then the client needs to
keep headers only for three last days (and keep dropping outdated ones), becoming a
\textit{partially light client}~\cite{light_cli}. In this example, the storage
required would be around 10MB. 
%
\begin{table}[tb]
\begin{center}
    \small
    \caption{The performance of \name certificate validation.}
    \label{tab:valid_perf}
    \begin{tabular}{c|rrrr}
        &\textbf{Min}  &   \textbf{Max}   &   \textbf{Avg.}  &  \textbf{Med.}   \\ \hline
  [\textit{ms}]&6.427 &   11.142 &    7.446 &    7.217 \\
    \end{tabular}
    \end{center}
\end{table}
\name clients validate obtained \name certificates (see
\autoref{sec:details:cert_valid}) by sanity and control
checks, and the validation of storage and code proofs.
To evaluate computation overhead, we executed
one hundred \name certificate validations and measured their execution time.
The obtained performance results are presented in
\autoref{tab:valid_perf}.  As shown, even with our unoptimized implementation,
the average and median values are about 7ms.
It
is relatively small when compared to latencies of TLS
handshakes (around 200ms on average~\cite{naylor2014cost}) and
should not be noticed by users~\cite{gribble1997system}.

\section{Discussion and Practical Considerations} 
\label{sec:practical}
\subsection{Limitations}
\label{sec:limitations}
Although \name provides multiple benefits it comes with some limitations.
First, despite designing \name with deployability in mind, we are aware that our
system introduces changes to the current ecosystem that may be seen as radical.
We demonstrate that the system is feasible and efficient even when built upon
existing tools and platforms, but given deployment experiences
with other TLS PKI enhancement we should not assume that \name will be deployed
right away as it is. We rather predict that \name would have to be deployed
incrementally, what we discuss in \autoref{pre:practical:deploy}.

Second, multiple aspects of \name (like economics, trust assumptions,
and throughput) are influenced by the underlying blockchain platform.  We evaluated
\name in an open environment to show that the certificate cost is reasonable
even in such a challenging deployment scenario.  However, we do not see any
reason why \name could not be deployed with a more efficient permissioned
blockchain~\cite{androulaki2018hyperledger} (e.g., run by a consortium of CAs,
browser vendors, and non-profit organizations).  Alternatively, \name could be
also implemented on the top of incoming permissioned networks, like Facebook's
Libra~\cite{libra}, which would eliminate the costs of launching and maintaining a new
platform.

Lastly, projects offering free
certificates rapidly increase the adoption of TLS~\cite{aertsen2017no}.  We
see it as a very positive trend, but we do not believe that \name can compete in
adoption with (free) DV certificates as costs of deploying \name may hinder
domains satisfied with the current protection level.  Instead, we see our scheme
as a more secure alternative addressed to security-savvy domains.  We argue that
in the face of abandoning EV certificates, such domains have no other choice
than deploying domain-validated certificates. 

\subsection{Incremental Deployment}
\label{pre:practical:deploy}
%
\name certificates cannot be introduced just by replacing X.509v3 certificates,
as legacy TLS clients would not be able to interpret them.   One approach is to
leverage an extension mechanism of X.509v3 certificates~\cite{rfc5280}.  This
mechanism allows to create a signed content that is part of a certificate.
However, a \name certificate cannot be efficiently implemented as an X.509v3
certificate extension, as \name certificates are dynamic and their state changes
frequently (the X.509v3 certificate would have needed to be re-issued for every
such an update).  One way to deploy \name in a
backward-compatible manner is to serve \name certificates through the OCSP
stapling extension~\cite{rfc6961}, which was designed to piggyback revocation
status messages on
the TLS handshake.  Then, supported clients would be able to process \name
certificates, while unsupported clients
would process X.509v3 certificates ignoring the unsupported extension.
Usually, incremental deployment and backward compatibility create downgrade
attack opportunities~\cite{11992}. We envision that the presented deployment
models could be extended by mechanisms like \textit{strict} modes~\cite{rfc6797}
or pinning~\cite{rfc7469} to avoid downgrade attacks on \name.

\subsection{Parameters}
\label{pre:practical:params}
\name introduces epoch whose value
constitutes trade-offs between efficiency, availability, and security.  With a
shorter epoch, the CA conducts public-key validations more frequently,
minimizing the attack window. On the other hand, shorter epoch requires more
frequent interactions between the CA and the \name contract incurring higher
cost and overhead. Epoch determines also how long CAs and domain servers can
remain unavailable, and what are tolerable blockchain confirmation times to avoid
forks (around 10 minutes in Ethereum). Taking it into account and following a
similar discussion on the short-lived certificates, we
envision that epoch values could be in practice set between several hours and a
few days.  Moreover, \name could support an epoch value per policy or even per
certificate what would allow domain owners to adjust their
certificates to unusual use cases (like backup certificates or certificates for
inaccessible subdomains). 
\name contracts as presented use the blockchain's notion of time by calling the
\textit{now()} function that returns the timestamp associated with the current
block.  Open blockchains are peer-to-peer networks
but
to guarantee only small clock desynchronizations it is part of the
consensus rules that block timestamps are consistent with the previous
timestamps and the time of the nodes. Moreover, Ethereum
nodes are ``advised'' to synchronize their clocks with an NTP server, thus
the time within the network is quite precise (up to a several
dozen seconds) and does not constrain epoch values significantly.

CAs use current block hashes to prove that their validation proofs are fresh.
However, the time between the public-key validation and the \textit{update()}
method execution (i.e., between the steps \textit{1a} and \textit{3b}
from \autoref{fig:update}) can be greater than one block. To tolerate
such delays, \name contracts 
check multiple previous
blocks (see \textit{update()} in \autoref{alg:contract}).  The number
of blocks checked, let us denote it as $N$, is variable and can be platform- and
situation-dependent.  For instance, it is reported~\cite{weber2017availability}
that almost 95\% of Ethereum transactions are added and propagated in about
300 seconds (the experiments were conducted in 2017), while of the time of
writing the
\url{ethgasstation.info} website reports 115 seconds as an average delay for
95\% of transactions. 
Since the average time of
a new block arriving is around 15
seconds, we can
conservatively assume that in most cases $N\geq300s/15s=20$.
Epoch bounds $N$, i.e., $N\leq \mathit{epoch}/15s$ in this setting;
however, it can be further bound since Ethereum contracts can access
up to 256 previous blocks, giving
$N\leq256$, what in practice means that a CA has to execute a update call
within around an hour (precisely, $256\times 15s = 3840s$) after a validation
proof was obtained.  Such a margin allows CAs to avoid and recover from
blockchain forks, since Ethereum transactions usually require 30 confirmation
blocks to be seen as ``secure''.
An important client-side parameter is \textit{maxStale} (see
\autoref{alg:validation}), denoting the maximum tolerable age
the last public-key validation.  A value of
this parameter should be greater than the epoch, and similar to epoch, it
introduces
a trade-off between security and efficiency. By following standardization of
similar parameters in TLS PKI~\cite{CABForum}, we envision that it
can be set between a day and several days.

\subsection{TLS 1.3}
\label{pre:practical:tls13}
TLS 1.3~\cite{rescorla2016transport} was recently approved as an Internet
standard.  Although a quick upgrade to TLS 1.3 is rather unlikely and should
not be expected~\cite{haztls13,whytls13}, the protocol introduces a few
important changes.  As mentioned before, TLS 1.3 removes the timestamp values
from ClientHello and ServerHello random fields. That change does not influence
\name design much as it does not rely on these fields.  A more important change is
the removal of the ServerKeyExchange message type and the introduction of a new
CertificateVerify message type instead. Fortunately, the semantics of the
CertificateVerify message type are similar. In particular, that message is
signed and the signature is computed over random fields supplied by a client and
a server in their hello messages, so \name can be
easily adjusted to be operational with TLS 1.3.

\subsection{Enhancements}
\label{pre:practical:policies}

\myparagraph{Multiple Public Keys}
For easy description, we presented \name in a setting where a domain has one
public key. It is a common practice that for operational reasons large domains
posses and use multiple key pairs.  \name can be easily extended to
handle this case. Namely, the \name contract, instead of a single key, is
initialized with multiple public keys. Then, for a validation proof submitted,
the contract verifies if it was signed by a private key corresponding to
one of the preloaded public keys.

\myparagraph{Validation Policies}
We demonstrate \name in a setting that improves the currently used and dominant
Internet-based domain validation, as the key continuity is measured by periodic
public-key validations conducted by multiple CAs.  \name with ``history-reach''
certificates enables to define and enforce almost arbitrary validation policies;
thus, we believe that other interesting \name contract code template(s) and
validation logic could be proposed.  For instance, policies could specify that
the ratio of validation errors (\textit{errNo}) is less than a certain threshold
(e.g., 10\%).  Policies could be also extended by parameters regarding
TLS connection properties like the desired security level of used cipher suites
(e.g., forward secrecy) or HTTPS pinning~\cite{rfc7469}.
The policy framework could be extended to support other applications and 
their features, like e-mail policies~\cite{rfc4408,rfc7489}.

\myparagraph{Policy Contract Management}
We envision that the policy contract would be managed by a standardization body
like the CA/Browser Forum. Such a body, like in today's PKI, could influence
decisions on which CAs can be added or removed from the trusted set or specify
and standardize available security policies.
In particular, the policy contract could store CA lists per browser vendor,
since vendors have different lists today.
With such a modification, certificates would have validation state per vendor.
We also see potential advantages of \name in terms of the on-blockchain CA keys
management.  Blockchain platforms, as reliable time sources, by design allow to
establish an order of events, that may be especially helpful in revocation of
compromised CAs. Such revocations are risky for the ecosystem, as they
invalidate all certificates issued by the revoked CA as a side effect.  \name
could be extended to remove this collateral damage from the TLS
PKI~\cite{PKISN}, by revoking CA with specifying the compromise time. In such a
case, a \name contract would ensure that it was created prior to the compromise
time; thus, is still valid.

\myparagraph{Notifications}
Blockchain platforms provide features that facilitate
smart-contract-empowered applications. For instance, Ethereum provides
asynchronous event notifications, which can be particularly helpful when
combined with \name.  With this mechanism, one could implement the contracts,
such that every policy update, \name contract creation, or public-key
validation error, would trigger a notification accessible by blockchain (light)
clients almost immediately.  Then, domain owners would be able to listen for
these notifications and in almost real-time detect all policy changes,
certificates created for their domain names, or their validation errors. That
could reduce the attack detection time radically.

\section{Related work}
\label{sec:related}
There are many systems that try to improve the security of digital certificates
and PKIs.  Short-lived certificates (SLCs)~\cite{rivest1998can} aim to remove
a revocation system~\cite{rfc5280,rfc6960} by designing certificates to be
valid for a few days.  With SLCs
domains have to conduct domain validation more frequently what mitigates
short-lived MitM attacks, but as SLCs are irrevocable it also increases attack windows associated
with compromised keys. Despite multiple tries to introduce them
to the TLS PKI, SLCs were abandoned~\cite{slc-ballot}.
EV certificates~\cite{EVguidelines} aim to
improve the security of domain-validated certificates.  To obtain an EV
certificate, an entity has to prove that it \textit{a)} controls the domain
name, and \textit{b)} is acting on behalf of the entity that is controlling the
domain name.  The latter check is manual and may require a face-to-face meeting;
thus the value proposition of EV certificates is better security.
 Security benefits of EV certificates were
questioned~\cite{jackson08fgo,EVvalue} and their support in main browsers may
be deprecated in a near future~\cite{evcab}.  
%
CAA~\cite{rfc6698} and DANE~\cite{rfc6844} are attempts to introduce domain expressiveness to
the TLS PKI ecosystem, by allowing domains to publish at DNS(SEC) their
simple security policies, like trusted CAs. Unfortunately, these schemes share
issues of DNS(SEC), are difficult to enforce, and their adoption rates
are low~\cite{szalachowski2017short,scheitle2018first}
Notary systems such as
Perspectives~\cite{wendlandt2008perspectives} and
Convergence~\cite{marlinspikeconvergence} use a multipath
probing to give better guarantees to a client about the certificate she has
obtained. They introduce trusted notary servers, that when requested by a client, check
and return their views of the server's public key. Then, the client can compare
her view with the notary's view and make a decision about the connection.
Unfortunately, notary systems have privacy issues, increase the latency of TLS
handshakes, introduce new highly available trusted parties~\cite{bates2014forced,merzdovnik2016whom}.
In \name public-key validations are automatically validated by the
smart contract code and recorded, providing  reliable statistics.

\begin{table*}[t!]
    \centering
    \caption{Comparison of the most related systems and \name.}
    \label{fig:compare}
    \includegraphics[width=0.975\textwidth]{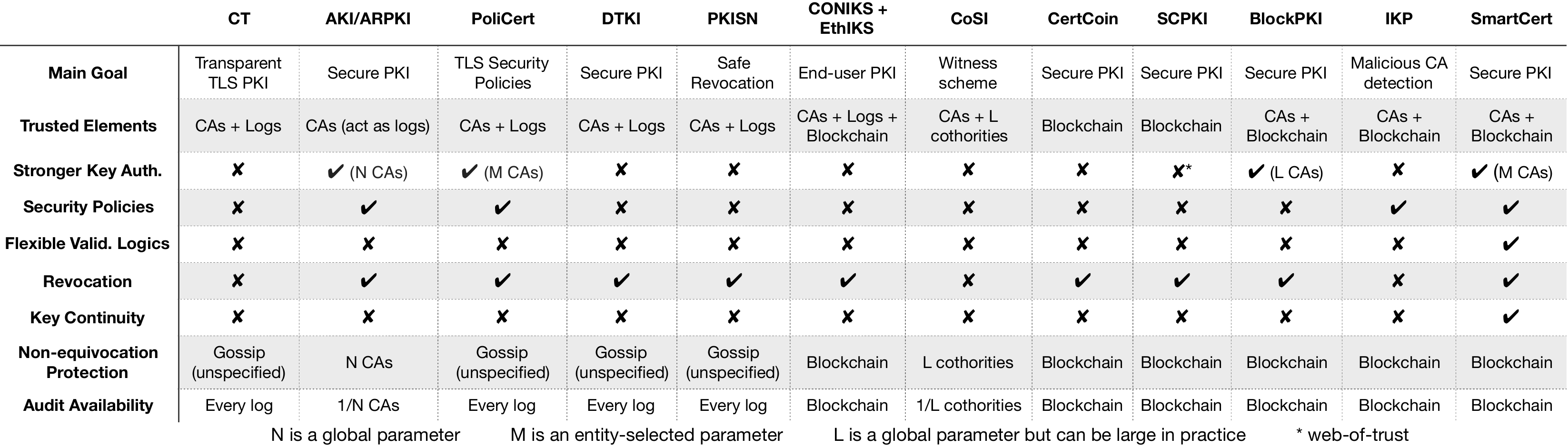}
    \vspace{-3pt}
\end{table*}

Sovereign Keys~\cite{eckersley2012sovereign} is a system logging
allowing domains to control (register or revoke) their
certificates and making them visible.
It is extended by
CT~\cite{rfc6962} which aims to
keep CA actions transparent. In CT, CAs submit all certificates (prior their
actual issuance) to a publicly verifiable log which for every submission returns a signed
promise that the certificate will be appended to the log's
database. 
A TLS client accepts a server's
certificate only if it is accompanied with the corresponding promise(s) from
trusted log(s).  As of the time of writing, CT is deployed with 32 public logs
operated by 10 different organizations.  Although CT is a significant
improvement of the TLS PKI, it has some drawbacks.  Logs do not run a consensus
protocol so each log's consistency has to be monitored separately
that requires external gossip
protocols~\cite{chuat2015efficient,nordberg2015gossiping} run by
clients.  Logs have to be highly available and from the initial deployment, this
seems to be challenging in
practice~\cite{ct_uptime,ct_uptime2}.  
Moreover, CT provides limited
security properties, aiming for certificate visible only, without asserting on its validation state. 

Many systems improve CT's security, efficiency, or add
novel functionalities. For instance, ARPKI~\cite{basin2014aar} (an
improvement over AKI~\cite{kim2013aki}) proposes a strongest-link PKI that
remains secure even if an adversary is able to compromise all trusted entities
but one.  Properties of ARPKI are formally verified; however, the system
introduces a static security parameter and requires significant changes to CAs.
ECT~\cite{ryan2014enhanced} enhances CT by introducing revocation and supporting
the e-mail communication, while DTKI~\cite{yu2016dtki} tries to relax
trust assumptions of the previous system, but assuming a trusted initial setup with
an honest log. The system is formalized; however, misses an implementation and
practical considerations (e.g., ECT and DTKI increase the latency of TLS
handshakes).  Another approach with a similar goal is
 a distributed witness framework CoSi~\cite{syta2016keeping} where
every seen certificate is asserted by a large group of authorities
(instead of one or few, as in CT).  The protocol scales due to a multi-signature
scheme deployed and its hierarchical network structure; however, such a design
requires coordination among authorities.  CONIKS~\cite{melara2015coniks} is an
attempt of providing key transparency to end users. The system relies on
publicly-verifiable logs and provides interesting features like users privacy.
PoliCert~\cite{szalachowski2014psf} is a log-based system that aims to better
support domain expressiveness. Domains in PoliCert can use public logs
to publish fine-grained policies describing their security preferences which then are
enforced by clients.  
All systems in that class share the limitations of CT. They require either trusted
validators or a gossip protocol run by clients (this aspect is usually
unspecified) and assume highly available logs. Moreover,
some of them increase the latency of TLS handshakes and violate
privacy by enforcing clients to contact logs, require new infrastructures,
or need significant changes to the TLS protocol and its PKI.  Unlike these
systems, \name achieves the transparency, non-equivocation, availability, and
censorship-resistance basing on the blockchain platform; thus, without
introducing a new infrastructure.  Moreover, in contrast to related work where
CAs are monitored by some other (new) trusted entities, CAs actions in \name are
monitored and validated by smart contracts themselves.

Up to our best knowledge, there is no work proposing a similar to
\name concept, the blockchain technology has been investigated in the context of
PKI recently. Namecoin~\cite{loibl2014namecoin} 
proposes a distributed namespace associated with public keys, what is improved
by Certcoin~\cite{fromknecht2014certcoin} which adds new functionalities (i.e.,
key revocation and
recovery).  Blockstack~\cite{ali2016blockstack} builds upon Namecoin, combining a distributed namespace and a
storage system.  Blockstack allows anyone to bind a name with a
controlled public key and it leverages blockchain properties to make such
bindings transparent, available, and equivocation resilient.  The platform is
fully implemented and deployed.
Wang et al.~\cite{wang2018blockchain} propose to implement CT over
blockchain adding a revocation functionality.  Their work requires an
application-specific blockchain, what may limit applicability and deployment of
the system.  It also abstracts from the current TLS PKI and 
and the complexity of real-world deployment.
SCPKI~\cite{al2017scpki} proposes a blockchain-based identity 
system employing the web-of-trust model~\cite{abdul1997pgp}. Interestingly,
SCPKI targeting end-user identities uses an external storage for storing
attribute data.  Patsonakis et al.~\cite{patsonakis2017towards} 
propose a blockchain-based PKI, whose main novelty  is a
constant size cryptographic accumulator within a smart contract storing and
managing certificates.  The work provides a formal analysis but misses an
implementation and evaluation.
BlockPKI~\cite{BlockPKI18} provides an automated blockchain-based framework
where multiple CA, synchronized via smart contracts,
efficiently issue certificates. The system does not support any domain policies
and relies on short-lived (irrevocable) certificates.
Blockchain platforms are also employed to facilitate
detection of misbehaving trusted parties.
EthIKS~\cite{bonneau2016ethiks} is a proposal for monitoring CONIKS logs using
Ethereum, where a special smart monitors the consistency of a log, replacing a required gossip protocol.  Another system is
IKP~\cite{Reischuk16:IKP}, where dedicated smart contracts incentivize
blockchain users for detecting fraudulent certificates.

In \autoref{fig:compare} we compare the most related work with \name.  For every
scheme we list its goal, trust assumptions, and important features like key
authentication, validation logics, policies, revocation, key continuity,
non-equivocation protection, and availability.

\section{Conclusions}
\label{sec:conclusions}
This paper proposes \name, a novel digital certificates framework.  By
leveraging the blockchain and smart contract technology, \name introduces
``dynamic'' certificates whose states can change but only according to the
encoded validation logic and security policies specified individually by domains.
Thanks to this design, \name provides properties and features that the previous
systems could not provide.  Certificates in \name self-enforce flexible security
policies, carry historical statistics of the conducted public-key validations,
the trust placed in CAs is minimized as their actions are transparent and
monitored by the code, and the system provides high availability, robustness,
and censorship resistance.  \name does not require major changes to the TLS PKI
and is combined with the TLS protocol.  The system is practical, does not
introduce significant overheads, and can be deployed incrementally, co-existing
with X.509v3 certificates.

We presented \name in the context of public-key validation; however, in the
future, we will investigate other validation policies. For instance, the \name
framework could be extended to support more sophisticated security policies
allowing domains to instruct the client about other attributes
of conducted connections (e.g., a domain's desired cipher suites).


\bibliographystyle{IEEEtranS}
\bibliography{ref,rfc}


%
\end{document}